\documentclass[aps]{revtex4}
\usepackage{epsfig}
\usepackage{color}
\usepackage{dcolumn,amsmath,amsthm,amscd,amsfonts,amssymb,graphics,graphicx,eucal}
\newcommand{\pd}{\partial}

\begin{document}

\date{\today}
\title{\textbf{\Large{Stable radiating gap solitons and their resonant interactions with dispersive waves in systems with parametric pump}}}
\author{Alexey V. Yulin$^{1}$, Leonardo R. Gorj\~{a}o$^{1}$, Kestutis Staliunas$^{2}$}
\affiliation{$^{1}$Centro de F\'{i}sica Te\'{o}rica e Computacional, Av. Prof. Gama Pinto, 2, 1649-003 Lisboa, Portugal \\
$^{2}$ICREA and Departament de F\'{i}sica i Enginyeria Nuclear, Universitat Polit\`{e}cnica de Catalunya, Colom 11, 08222 Barcelona, Spain }

\begin{abstract}
We study the formation of gap solitons in the presence of parametric pump. It is shown that parametric pump can stabilize stationary solitons continuously emitting dispersive waves. The resonant interactions of the radiation and the solitons are studied and it is shown that the solitons can be effectively controlled by the radiation. In particular it is shown that the solitons can collide or to get pinned to inhomogeneities due to the interactions mediated by the resonant radiation.
\end{abstract}

\maketitle


\section{Introduction}

The interaction of solitons is a key issue in the theory of conservative solitons. One of the strict definitions of the solitons is that they collide elastically, i.e. recover initial velocities after collision, and experience no radiation in the collisions \cite{Zabusky,Ablowitz,Newell}.

The collisions of the solitons in a broad sense of the term soliton (dissipative solitons \cite{Akhmediev} or cavity solitons in optics \cite{Chen,Kestas_book}) are not elastic. Sometimes the solitons can merge in a collision, sometimes they can form bound soliton pairs \cite{Akhmediev,Gordon,Reynaud,Krolikowski,Stegeman}. Solitons can be utilized for solution of various applied problems \cite{SolApp}, in some cases inter-soliton interaction is a drawback e.g. in the lines of information communication based on solitons \cite{Hasegawa}, sometimes the interaction can be useful, e.g. for lasers \cite{Grelu,Turitsyn1}, supercontinuum generation \cite{Gorbach,Turitsyn2}. Typically the solitons interact locally, due to overlap of their shapes in space \cite{Gordon,Schapers,Ultanir}, although the global coupling was also reported \cite{kestas,kestas2}.

Recently the dispersive wave mediated inter-soliton interactions has been discovered and reported in a number of papers \cite{Smith,Rotschild,Vladimirov,Jang,OE_yulin}. It is known that solitons can be in resonance with the free propagating modes of the medium and then they radiate on their own frequency. This radiation often referred as Cherenkov radiation \cite{Cheren1,Cheren2,Cheren3} affects the dynamics of the solitons and can lead to long-range interaction between the solitons. The inter-soliton interaction can also be mediated by the radiation appearing because of the Hopf bifurcation leading to the formation of oscillating solitons \cite{Elphick,Vladimirov,OE_Skryabin}.
Different scenarios have been predicted, where the radiation circulating between the solitons either attracts solitons and cause them to annihilate, or result in the formation of pairs of bound solitons.

In the present paper we focus on the radiation of the dispersive waves by band-gap solitons and on the inter-soliton interactions caused by the radiation. The bandgap solitons have their frequency lying in the gap of spectrum of the linear problem \cite{Gap1,Gap2}. The bandgap solitons do not radiate via Cherenkov mechanism due to the absence of the resonant modes in the band-gap, however they can experience oscillatory instability \cite{Pelinovsky_stab_gap} resulting in the energy transfer from the soliton to delocalized dispersive waves. As interpreted recently \cite{Gaizauskas} two excitation quanta of the solitons can annihilate and reappear one in the upper, other in the lower band on a top and bottom of the bandgap. When the frequency of the soliton is around the middle or in the lower half of the bandgap such radiation process becomes very efficient as the energy and the momentum conservation holds. This is also very desirable for the purposes of this article, as the frequency of the radiation is very distinct from that of the soliton.

In conservative systems the emitting solitons can exist only for a limited time because of radiative losses. It would be very desirable to design a configuration, where the gap solitons would permanently emit the radiation: this would open new possibilities of soliton interaction schemes.  Attempts to restore the decaying solitons by a linear gain are also usually not fruitful: this amplification  feeds not only the solitons, but also the delocalized radiation, which destroys the soliton state.

The solution which we propose in the present article is to use a highly frequency selective parametric gain to amplify the solitons, and thus to prevent their decay. On the other hand, if the dispersive radiation is of different frequency than that of the soliton, then that radiation will be not affected by the parametric gain. This would enable to have permanently radiating but stable solitons.

The aim of the article is to substantialize the above outlined idea. We consider the model, where the field is decomposed into two interacting waves. These components depending on the physical system can be - either forward-backward propagating waves (for fiber Bragg solitons \cite{Gap1}), of left-right propagating waves (spatial solitons \cite{Gap2,Gap3,Gap4,Gap5}). The mathematical description, in terms of two coupled equations is, meanwhile is well established \cite{Gap2} and universal, therefore we do not consider the details of the physical system.

The structure of the paper is the following. First we consider the mathematical model of the considered system. Then we study the stability of the solitary solution of the system (gap solitons) and show that the parametric pump can stabilize solitons and make them stationary. In the following section the resonant interactions between solitons and dispersive waves are studied and the radiation mediated inter-soliton interactions are considered.  Next we show that the reflection of the soliton emitted waves on an inhomogeneity can results in pinning of the soliton to that inhomogeneity. Finally in the conclusion we briefly summarize the main results of the paper.

\section{The model}

\begin{figure}[h]
\centering \epsfig{file=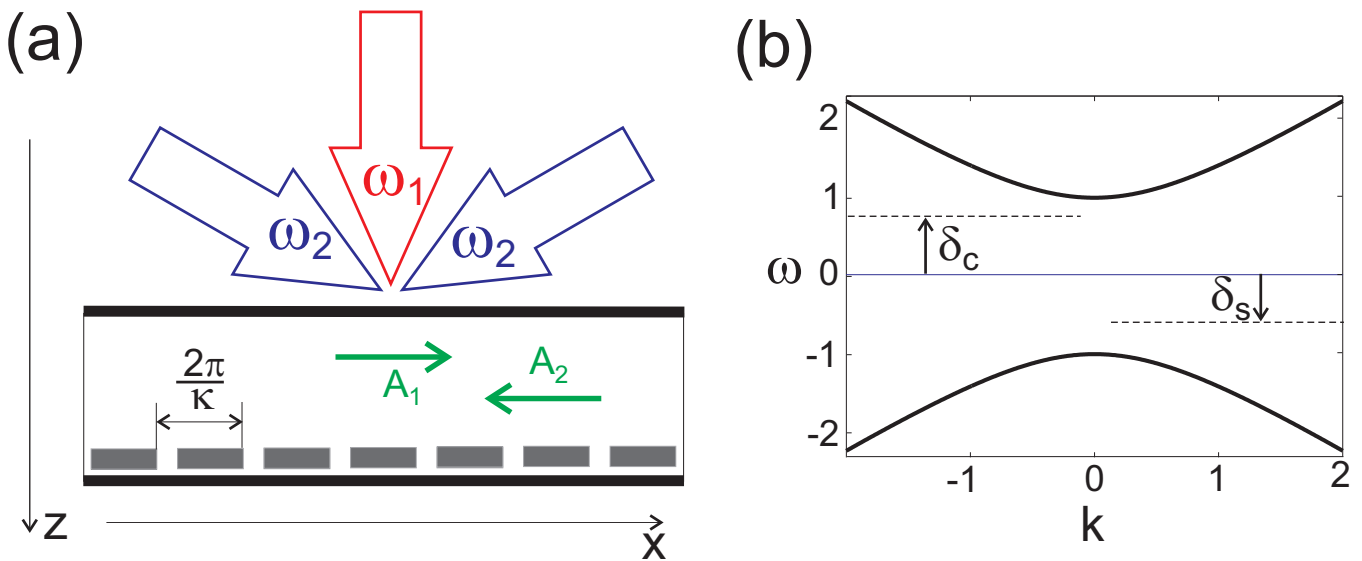, width=\textwidth}
\caption{The system considered in the paper is shown schematically in panel (a). Thick lines are the mirrors forming the cavity, grey rectangles are the periodical inhomogeneities build-in the cavity. The counter-propagating resonant waves are shown by green arrows and marked as $A_1$ and $A_2$. In the bottom of the panel $x$-axis is shown. The external pump providing the parametric gain is shown by the blue and red arrows, the projection of the wave vector corresponding to the light with $\omega_2$ (blue lines) on $x$ is equal to the half of the lattice constant $\kappa/2$. The dispersion characteristics in the vicinity of the gap is shown in panel (b) by thick black lines. The thin blue line marks the centre of the gap. The detuning of the resonance frequency of the parametric pump from the centre of the gap is marked as $\delta_c$. The solitons can be characterized by the detuning of their frequency from the centre of the gap, this detuning is shown as $\delta_s$. Let us note that the detunings $\delta_c$ and $\delta_s$ have opposite signs, so the soliton is in the resonance with the parametric pump if $\delta_c+\delta_s=0$.}
\label{fig1}
\end{figure}

We consider evolution of light in a system schematically shown in panel (a) of Fig.~\ref{fig1}. This is a cavity formed by the mirrors of high reflectivity at the frequency of solitons. The cavity can support modes propagating to the right and to the left. Inside the cavity there is a periodical inhomogeneities shown as grey rectangles with the period $L=2\pi /\kappa$ where $\kappa$ is the lattice constant. Due to this inhomogeneity the waves with wavectors $q = \kappa/2$ propagating in opposite direction can effectively interact with each other (Bragg scattering). This results in a gap in the dispersion characteristics $\omega(k)$. The dispersion in the vicinity of the gap is shown in panel (b) of Fig.~\ref{fig1}. Such linear cavities have been experimentally realized and reported \cite{pescus}

We also assume that the cavity has Kerr nonlinearity and that is pumped by two laser beams with the frequencies $\omega_1$ and $\omega_2$ such that $\omega_2-\omega_1 \approx 2 \omega_0$, where $\omega_0$ is given by the dispersion relation of the guided modes $\omega_0=\omega(q)$.

The only novelty which differs our system from the similar systems considered in \cite{Gap2,Gap5} is the external pump at the frequencies $\omega_1$ and $\omega_2$. Let us briefly show how this holding beams lead to the parametric pump for the cavity modes. Assuming that the grating is relatively shallow and the nonlinear effects are weak we can use coupled wave approach describing the field $E$ in terms of the slow varying amplitudes of the counter-propagating waves, $E=A_1(t, x) e^{iqx -i\omega_0 t}+A_2(t, x) e^{-iqx -i\omega_0 t}+cc$, where $\omega(k)$ is the dispersion of the guided modes of the cavity. In the case of shallow grating and weak nonlinear effects the holding beam can be represented as
$$E_p=B_1(z) e^{i k_1 x - i \omega_1 t} +B_2(z) e^{i k_1 x + i \omega_1 t} + C(z) e^{- i \omega_2 t} +cc$$ where $B_1$, $B_2$ and $C$ are the functions found from the solution of the linear problem for the reflection and transmission of the pumping beams on the cavity, $k_1$ is the projection of the wavevector of the beam with frequency $\omega_1$ on the horizontal axis.

We assume that the distribution of the pump field is known (in the leading approximation order). Now we can write the term describing the polarization caused by the Kerr nonlinearity
\begin{align}
P=\chi \left( \xi A_1(t, x) e^{iqx -i\omega_0 t}+\xi A_2(t, x) e^{-iqx -i\omega_0 t } + \right. \nonumber \\
 + B_1(z) e^{i k_1 x - i \omega_1 t} +B_2(z) e^{i k_1 x + i \omega_1 t} + \\
\left. + C(z) e^{- i \omega_2 t} +c.c.  \right)^3 \nonumber
\end{align}
here we denoted $q=\kappa/2$ and $\omega_0=\omega(q)$, $\xi(z)$ defines the structure of the guided mode (along $z$) and $\chi(z)$ is the nonlinear coefficient (function of $z$).

Under the assumption that $\omega_2-\omega_1 \approx 2 \omega_0$ and $k_1 = 2q$ the only term
\begin{align}
3\xi(z)e^{iqx-i\omega_0 } \left( |\xi A_1|^2+2|\xi A_2|^2+ \right. \nonumber \\
\left.+2|B_1|^2+ 2|B_2|^2+2|C|^2 \right)A_1+ \label{res_to_emod} \\
+2\chi(z) \xi(z)^{*} e^{-i(\omega_2-\omega_1-\omega_0)t }A_1^{*} B_2^{*} C  e^{i(k_1-q)x}  \nonumber
\end{align}
can be in resonance with the guided mode $A_1 e^{iqx -i\omega_0 t}$.
The first two lines of (\ref{res_to_emod}) will result in the self-modulation term $\sim \left( |A_1|^2+2|A_2|^2 \right) A_1$ and the constant frequency shift in the equation for slow varying amplitude $A_1$. The last line will produce a parametric term $\sim B_2^{*} C e^{-i(\omega_2-\omega_1-\omega_0)t }A_1^{*}$. Note, that the coefficient $B_2^{*} C$ can be set to be real without loss of generality.

Analogously one can derive the equation for $A_2$. Neglecting the dynamics of the holding beams, we finally arrive to the equation governing the dynamics of the light in the cavity which in dimensionless variables reads:
\begin{align}
\pd_t A_1 = -\pd_x A_1 -i(|A_1|^2 + 2|A_2|^2)A_1 ~+\nonumber\\ \quad i A_2 + i\delta_d A_1+ i\mu A_1^{*}e^{-2i\delta_c t}  - \gamma A_1 \label{e3}+f(t, x)\\
\pd_t A_2 = ~~\pd_x A_2 -i(|A_2|^2 + 2|A_1|^2)A_2 ~+\nonumber\\ \quad i A_1 + i\delta_d A_2 + i\mu A_2^{*} e^{-2i\delta_c t} - \gamma A_2,\label{e4}
\end{align}
where $\mu$ is the parametric gain coefficient, $\delta_c=(\omega_2-\omega_1-2\omega_0 )/\Delta$ is the detuning of the resonant frequency of the parametric pump from the centre of the gap normalized on the gap width $\Delta$ (see panel (b) of Fig.~\ref{fig1}) and $\gamma$ is the coefficient of the linear losses in the system. The dependence of $f(x, t)$ accounts for the driving force (the amplitude of a probe beam pumping the system from the top, not shown in Fig.~\ref{fig1}). The parameter $\delta_d(t, x)$ accounts for possible dependency of the position of the centre of the gap on space and time. For most of the paper $\delta_d=0$ apart from the section devoted to pinning of solitons on spatial inhomogeneities. The time is normalized on the gap width $\Delta$, the space coordinate is normalized on $\frac{2\omega(\kappa/2)}{\kappa\Delta}$. For numerical simulations we used absorbing boundary conditions in order to avoid wave reflections from the boundaries. For the rest of the paper apart from the last section we assume that $f=0$ meaning that we do not have a direct resonant pump in the system. Let us mention that the explicit dependence on time can be removed from the equations by changing the variables $A \rightarrow Ae^{-\delta_c t}$.

For the case without losses $\gamma=0$ and without parametric pump $\mu=0$, the dispersion is very simple
\begin{align}\label{dispersion}
\delta_{\pm}=\pm \sqrt{k^2 +1},
\end{align}
where $\pm$ stands for the upper and the lower dispersion branch correspondingly. The solitons with the frequency lying in between these branches are known to exist, and are named gap solitons. We study how these solitons can be stabilized against radiative instability and investigate how the solitons can interact with quasi-linear propagating waves.

The solitons can be characterized by their detuning from the centre of the gap $\delta_s$ (see panel (b) of Fig.~\ref{fig1}), the amplitude and the width of the solitons are functions of the detuning $\delta_s$. The analytical solution of the zero velocity gap soliton has the form
\begin{align}
A_1=\sqrt{\frac{1-\delta_s^2}{3}}\text{sech}\left(\frac{x}{1-\delta_s^2} + \frac{i}{2\arccos(\delta_s)}\right), \nonumber \\ A_2=A_1^{*}. \label{ansatz}
\end{align}

\begin{figure}[h]
\centering \epsfig{file=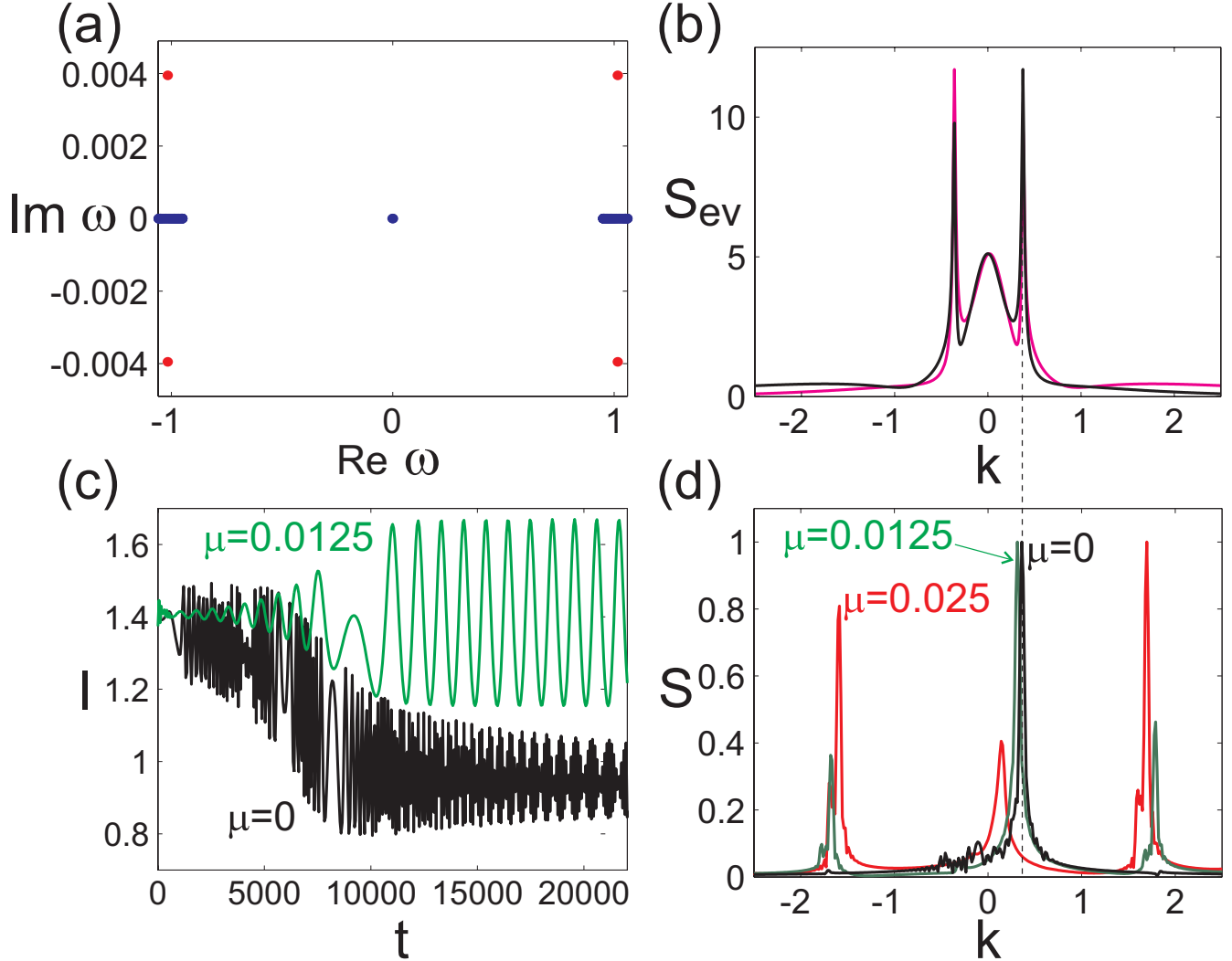, width=\textwidth}
\caption{Panel (a) shows the spectrum for the soliton solution with $\delta_s=-0.05$. Four red dots show the quartet of the eigenfrequencies generating the instability. The spectrum of the eigenvector corresponding to the unstable mode is shown in panel (b), the black line corresponds to $A_1$ field and the purple line to $A_2$ field. Panel (c) shows the dynamics of the amplitude $I=max(|A_1|^2+|A_2|^2)$ of the soliton for the conservative case in the absence of the parametric pump (black line) and for the dissipative case with parametric pump $\mu=0.0125$ (green line). The spectra calculated for the conservative case (at $t=500$) and for the cases with parametric pumps with $\mu=0.0125$ and $\mu=0.025$ (at $t=20000$) are shown in panel (d). For all cases with parametric pump  the loss is $\gamma=0.005$ and the detuning is $\delta_c=0.05$. The spectra shown in panel (d) are calculated in the interval $x=[10, 150]$.}
\label{fig:instab}
\end{figure}

The stability of such solitons depends on $\delta_s$, \cite{Pelinovsky_stab_gap}, i.e. on their position inside the bandgap. Here we remind the results of the spectral analysis of the stability of the soliton. We linearize \eqref{e3}-\eqref{e4} around the soliton solution \eqref{ansatz} and solve the corresponding spectral problem governing the stability of the solitons. The solitons are unstable for negative soliton frequency detuning, in other words the solitons are get destabilized when their intensity exceeds a threshold limit. The instability is generated when two imaginary eigenvalues belonging to the discrete spectrum collide with the continuum and produce a quartet of eigenvalues. A typical spectrum of the unstable soliton is shown in panel (a) of Fig.~\ref{fig:instab}. Panel (b) shows the spatial spectrum of the eigenvector associated with the instability. The eigenvector can be considered consisting of two parts: the field localized in the core of the soliton and wide wings of weakly localized radiation. The instability leads to the oscillation of the soliton and to the emission of the waves with the frequency $Im~ \omega$ and the wavenumbers $k$ corresponding to the narrow spectral lines in panel (b).

\begin{figure}[h]
\centering \epsfig{file=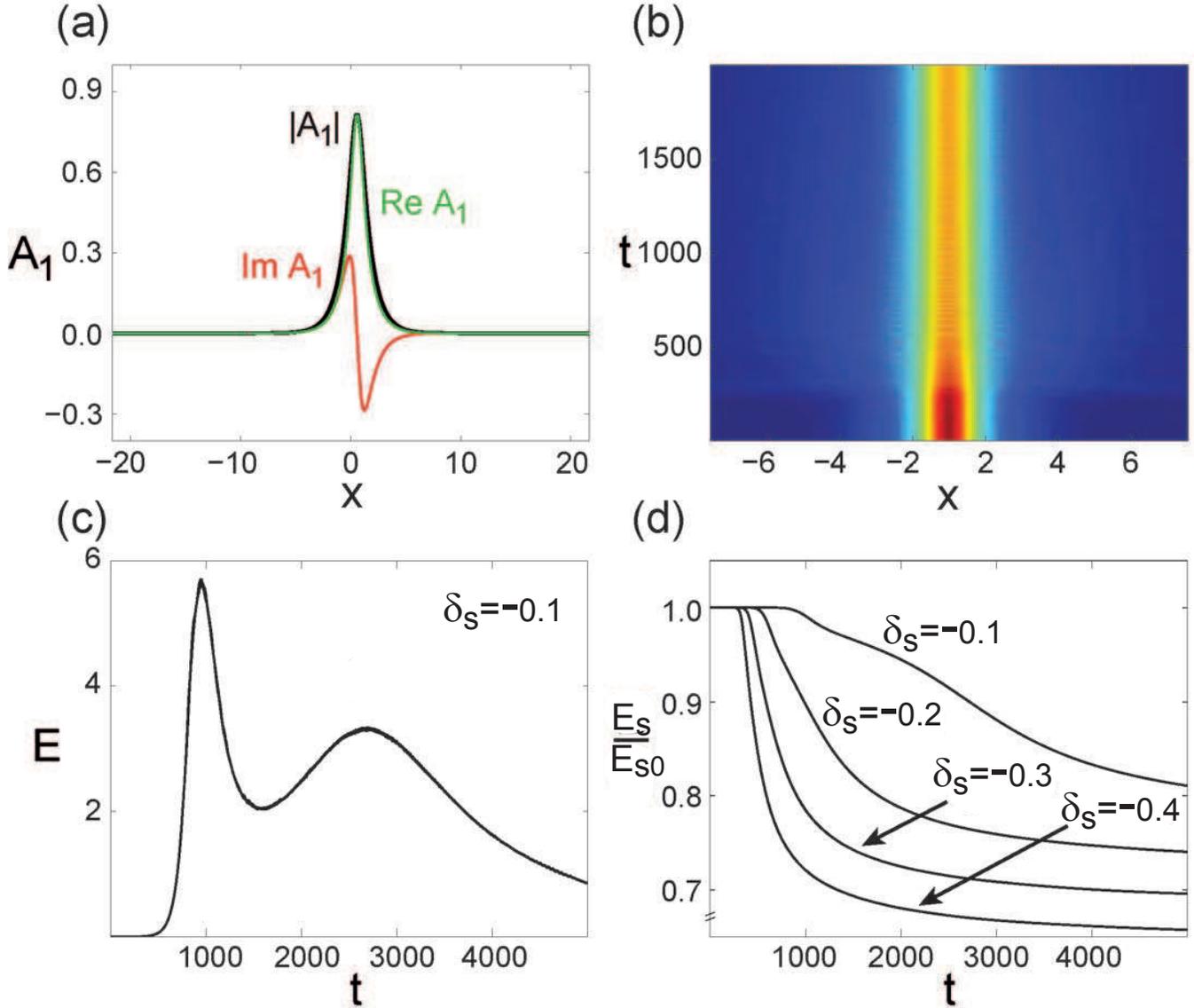, width=\textwidth}
\caption{Panel (a) shows an exact conservative soliton solution for $\delta_s=0$. The dynamics of an unstable soliton with $\delta_s =-0.3$ perturbed by neak noise is shown in panel (b). Panel (c) shows the temporal variation of energy $E(t)=\int |A_1|^2+|A_2|^2 dx $ in the interval x=[10,150] right of the soliton, the soliton parameter is $\delta_s=-0.1$.  Panel (d) displays the dependencies of energy $E_s$ of unstable solitons with initial $\delta_s= -0.1, -0.2, -0.3$ and $-0.4$ on time. Energies are normalized to the initial soliton energy $E_{s0}=E_s(t=0)$.}
\label{fig:Picture1}
\end{figure}

The results of the solution of the corresponding spectral problem are confirmed by the direct numerical simulations of the initial equations \eqref{e3}-\eqref{e4}. As the initial conditions we took the analytical soliton solution \eqref{ansatz} (panel (a) of Fig.~\ref{fig:Picture1}) perturbed by weak noise. The results of numerical simulations are presented in panels (b)-(d) of Fig.~\ref{fig:Picture1}.

In panel (b) of Fig.~\ref{fig:Picture1} the evolution of the soliton is shown. After some period of time the instability develops and radiation starts to emerge from the soliton structure. Take notice that the soliton starts to oscillate. The intensity of the radiation grows at first because the developing instability increases the emission rate of the dispersive waves from the soliton.
To see this we can watch how the energy of radiation is changing in the certain interval (of fixed length) outside the soliton. The radiation can come there only from the soliton, that is why initially the energy of this radiation is practically zero and stays very low until the instability develops and the emitted waves reach the control area, see panel (c) of Fig.~\ref{fig:Picture1} showing this.

But the propagating waves, \emph{wings}, take some of the solitons energy, so the intensity of the soliton goes down, see panel (c) of Fig.~\ref{fig:instab} where the dependence of the soliton energy on time is shown. The decrease of soliton intensity suppresses the emission and the soliton gets stabilized at some intensity smaller than the critical one. If we measure the radiation energy in the control area outside the soliton we will see that it will go down because very little radiation is now coming from the soliton, see panel (c) of  Fig.~\ref{fig:Picture1}.

To prove that the radiated field appears because of the instability we can calculate the spectrum of the field emitted by the soliton and compare it with the spectrum of the unstable mode. To make the picture more instructive we remove the contribution from the soliton calculating the spectrum in some area situated on the right of the soliton relatively far from it. This will be the spectrum of the emitted waves propagating to the right. Of course this procedure filters out the emitted wave propagating to the left. The spectrum calculated in the window $x=[10, 150]$ is shown in panel (d) of Fig.~\ref{fig:instab}.  In the same panel the spectrum of the unstable eigenmode is shown. It is seen that the positions of the spectral lines of the emitted field practically coincide with the position of the spectral line of eigenvector governing the unstable mode.

 We did numerical simulations for the solitons with different initial $\delta_s$ and demonstrated that more intense solitons decay faster then the low intensity solitons. Panel (d) of Fig.~\ref{fig:Picture1} show the dynamics of the normalized energy of the soliton on time. It is seen that the intense solitons start to decay sooner and decay faster then the solitons of low intensities.  We have also checked by direct numerical simulations that for the positive $\delta_s$ the solitons are stable and can propagate for very long times without any changes.

The next issue we are going to address is if we can pump the gap solitons and compensate for their radiative losses. It would open a possibility to obtain stable stationary gap solitons constantly emitting dispersive waves. To prevent the destruction of the solitons by growing linear waves we need to provide a pump delivering energy to solitons but not to the linear waves. This can be achieved by the frequency sensitive parametric pump with the resonant frequency lying in the gap. Since there are no linear waves with frequency lying in the gap, the pump will not amplify the linear waves.  On contrary, the soliton frequency lyes in the gap and  thus the parametric gain can pump the soliton.

\begin{figure}[h]
\centering \epsfig{file=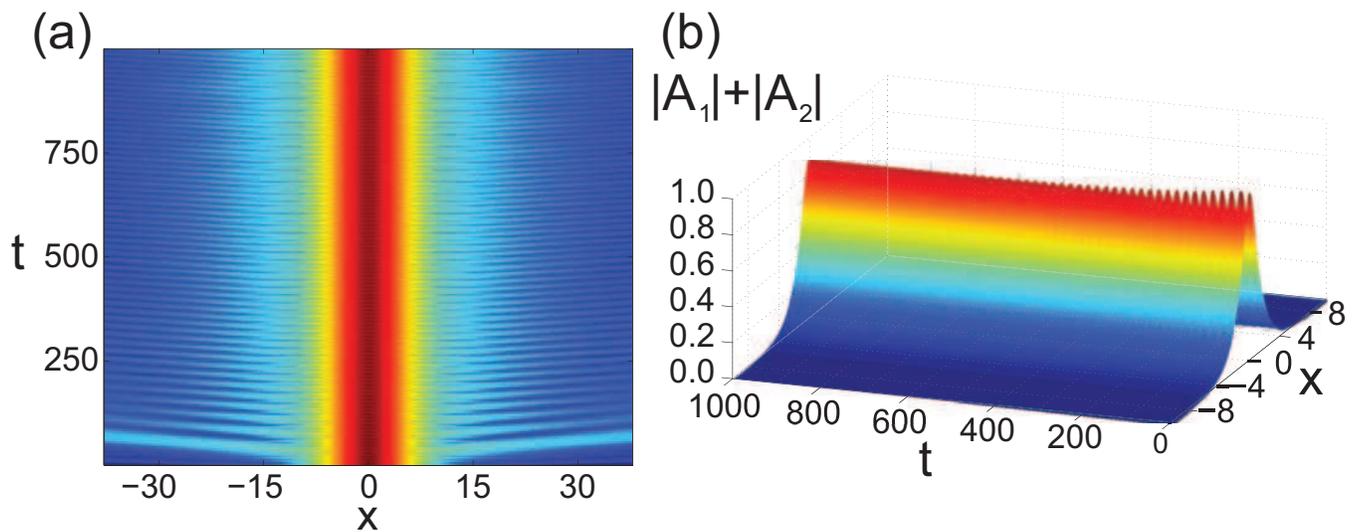, width=\textwidth}
\caption{Panel (a) shows a soliton with positive $\delta_s=0.2$ which is set to radiating behavior by fixing the parametric pump frequency below the stability threshold, at $\delta_c=0.25$ with $\mu=0.025$. A permanent oscillatory dynamics takes place, emitting waves and keeping in balance the inbound at outbound energy of the soliton. Panel (b) shows the evolution of a soliton with negative $\delta_s=- 0.05$ with the parametric pump frequency being in resonance with stable solitons $\delta_c=-0.1$. The stabilization of the soliton is clearly seen, no energy is emitted from the soliton after the stabilization. The pump intensity is $\mu=0.025$ and the linear losses in the system is $\gamma=0.005$. }
\label{fig:EinPump}
\end{figure}

To obtain the dispersive wave emitting solitons we set the frequency of the pump so that it is in the resonance with an unstable soliton. In particular we choose $\delta_c=0.25$ and $\mu=0.025$ for the pump and as the initial condition we took a soliton solution with $\delta_s=0.2$.  We also introduced very small linear losses $\gamma=0.005$. In panel (a) of Fig.\ref{fig:EinPump} one can see the formation of the radiation wings and the stabilization of the soliton developing into a oscillating localized structure. The soliton energy is maintained in a narrow window of values, thus finding a balance between the losses caused by the emission and gain produced by the parametric pump. The presence of small linear losses happens to be essential for the stabilization of the solitons.

If the parametric pump is tuned to be in resonance with stable solitons then a non-oscillating state can form. Panel (b) shows the case when the parametric pump has frequency $\delta_c=-0.1$ and amplitude $\mu=0.025$. The initial conditions were chosen in the form of the unstable soliton given by formula (\ref{ansatz}) with  $\delta_s=- 0.05$. It is seen that after some oscillations the soliton transforms into a non-oscillating stable solitary state.  Of course direct numerical simulations cannot give an ultimate prove that the soliton is stable but we found that the solitons can exist for very long times without any noticeable changes in their shape or velocity. This gives very strong indication that the solitons are stable.

In panel (c) of Fig.~\ref{fig:instab} the evolution of the top intensity of the soliton in time is shown for the case with parametric pump and linear losses. It is seen that eventually an oscillating soliton state is established. The spectrum of the field calculated from the finite area on the right of the soliton is shown in panel (d). One can see that  for relatively low pump intensity $\mu=0.0125$ the spectral lines of the stationary field practically coincide with the corresponding lines of the spectrum of the unstable eigenmode. However for stronger pump $\mu=0.025$ the lines shifts to the left. At the same time the sidebands become much more pronounced.

We can conclude that the gap solitons in the presence of parametric pump can act as continous sources of the dispersive waves. In this light it is interesting to consider how the dispersive radiation can affect the gap solitons.

\section{Resonant scattering of the dispersive waves on gap solitons}

In the previous section we have shown that the parametric pump can stabilize radiating gap solitons. Now our purpose is to address the interaction between gap solitons and the linear waves. We start with the scattering of the quasi-linear waves on the conservative solitons without parametric pump. We will be looking for the solition in the form $A_{1, 2}=u_1+A_{s1, s2}+A_{p1, p2}$,  where $A_{p1, p2}=A_{p10, p20}e^{ik_i x -i\delta_i t}$ is the incident wave with the wave vector $k_i$ and the frequency $\delta_i=\delta(k_i)$. The linearized equation for the weak scattered waves $u_{1, 2}$ reads
\begin{align}
\pd_t u_1 + \pd_x u_1 -i A_2+2i(|A_{s1}|^2 + |A_{s2}|^2)\cdot u_1 + \nonumber \\
+2iA_{s1}A_{s2}^{*}\cdot u_2 + iA_{s1}^2\cdot u_1^{*} + i2A_{s1}A_{s2}\cdot u_2^{*}= \nonumber\\
-2i(|A_{s1}|^2 - |A_{s2}|^2) A_{p1} - 2iA_{s1}A_{s2}^{*} A_{p2} - \label{lin_eq1} \\
- iA_{s1}^2 A_{p1}^{*} - i2A_{s1}A_{s2}A_{p2}^{*}. \nonumber \\
\pd_t u_2 + \pd_x u_2 -i A_1+2i(|A_{s2}|^2 + |A_{s1}|^2)\cdot u_2 + \nonumber \\
+2iA_{s2}A_{s1}^{*}\cdot u_1 + iA_{s2}^2\cdot u_2^{*} + i2A_{s2}A_{s1}\cdot u_1^{*}= \nonumber\\
-2i(|A_{s2}|^2 - |A_{s1}|^2) A_{p2} - 2iA_{s2}A_{s1}^{*} A_{p1} - \label{lin_eq2} \\
- iA_{s2}^2 A_{p2}^{*} - i2A_{s2}A_{s1}A_{p1}^{*}. \nonumber
\end{align}

Separating real and imaginary parts of $u$ we can obtain a linear equation with the right hand side $\hat L \vec u= \vec f$. The expressions for $\hat L$ and $\vec f$ can easily be obtained however the formulas are quite long and we omit them for sake of brevity. The important thing is that the operator $\hat L$ has zero eigenvalues. The equations (\ref{lin_eq1})-(\ref{lin_eq2}) have stationary solution only if $\vec f$ is orthogonal to the eigenvectors of the adjoint operator $\hat L^{+}$ corresponding to zero eigenvalues of the operator. Otherwise the resonance will take place and solution will contain radiation tails of low intensity but with lengths growing with time.

\begin{figure}[h]
\centering \epsfig{file=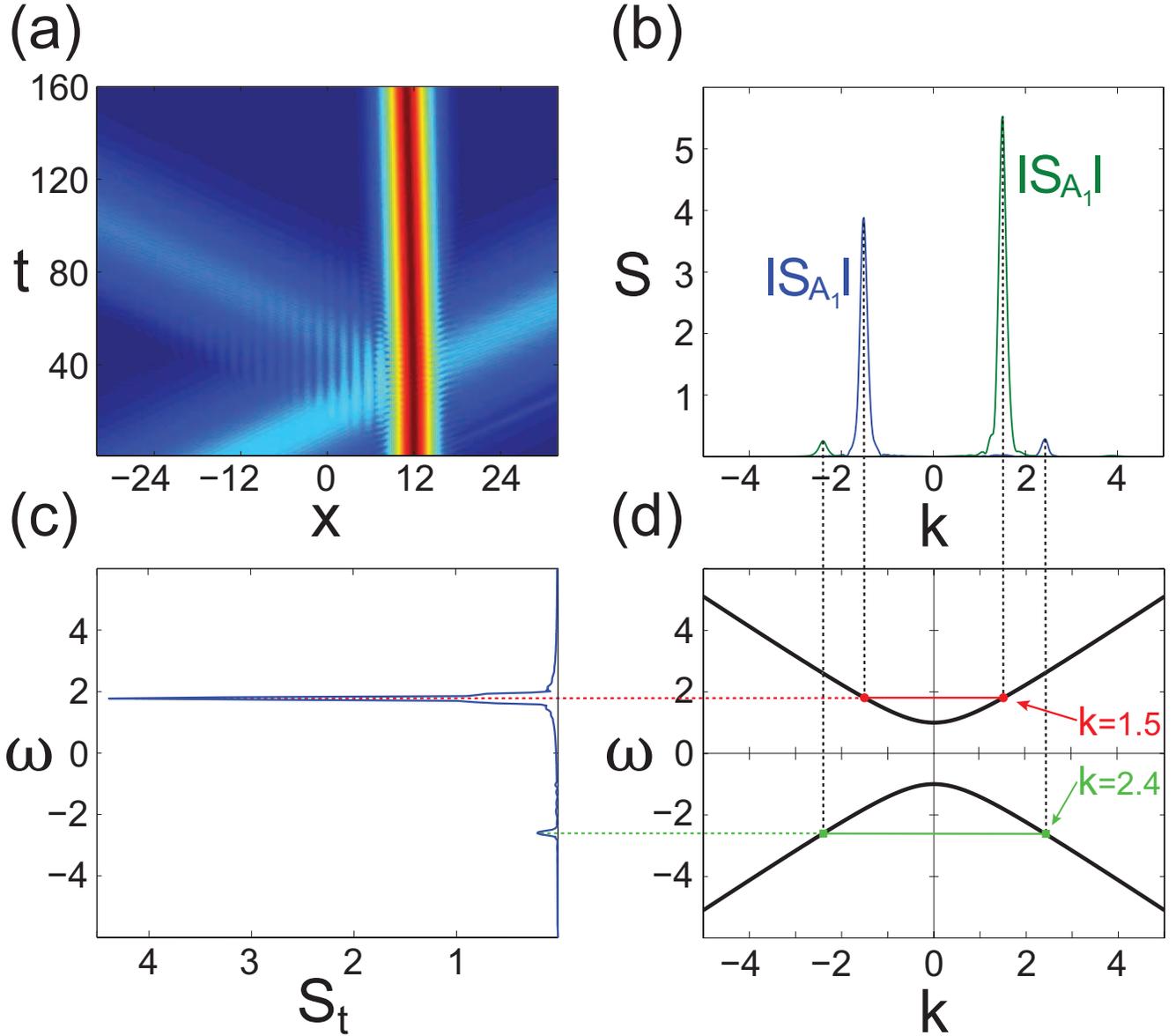, width=\textwidth}
\caption{Panel (a) shows a soliton at $\delta_s=0.4$ interacting with a propagating low amplitude plane wave with $k=1.5$, one can see the transmitted and reflected waves. In panel (b) the spectra of the reflected and transmitted waves at $t=160$ are plotted, displaying a peak at $k=-1.5$, which is merely the reflected wave with same $k$, and another peak at $k\simeq \pm 2.4$. Panel (c) shows the temporal spectrum $S_t$ of the transmitted waves. In panel (d) the dispersion law given by (\ref{dispersion}) are shown by the black curves, the graphical solutions of the resonance conditions (\ref{res1})-(\ref{res2}) are shown as the crossings of the dispersion characteristics with the green and red lines.  }
\label{fig:EPS/Fig1}
\end{figure}

Since the soliton solution $A_{s1, s2}$ is localized, the eigenvectors belonging to the continuum of the operator $\hat L$ have asymptotic in the form of plane waves  and so can be characterized by a wavevector $k$ of their tails at $x\rightarrow \pm\infty $. The frequencies of such eigenmodes are given by the dispersion $\delta_{\pm}(k)$. The condition of the resonance is that the spectral representation of $\vec f$ contains harmonics having the same wave vector and frequency as an eigenfunction of the operator $\hat L$ corresponding to a zero eigenvalue. We skip the full length derivation which is completely analogous to the derivation performed in \cite{yulin_OL_res_cond,yulin_PRE_scatt} and write down the final result. The resonant condition of four-wave mixing of the solitons with linear waves  gives us
\begin{align}
\delta_r = 2\delta_s - \delta_i \label{res1} \\
\delta_r=\delta_i \label{res2}
\end{align}
where $\delta_r=\delta_{\pm}(k_r)$ is the emerging resonant wave frequency and $k_r$ is the wave vector of the resonantly scattered wave. The condition is valid for resting soliton but can easily be reformulated for moving solitons. We note that the dispersion relation \eqref{dispersion} has two branches, thus four different modes with two different frequencies can be involved in the process. The said is illustrated in Fig.~\ref{fig:EPS/Fig1} where panel (a) shows the evolution of the soliton (the bright vertical stripe) and of the dispersive waves envelope. One can see that the collision of the dispersive pulse with the soliton results in partial reflection of the pulse on the soliton.

We have calculated the spatial spectrum of the radiation in the intervals on the right (the green curve in panel (b) of Fig.~\ref{fig:EPS/Fig1}) and on the left (the blue line in panel (b)) from the soliton after the interaction, when the reflected and transmitted dispersive waves and the soliton became well resolved in space again. One can see that the modes with four different $k$ are excited. The temporal spectrum of the transmitted radiation is shown in panel (c) of Fig.~\ref{fig:EPS/Fig1}. The resonance conditions (\ref{res1})-(\ref{res2}) are graphically illustrated in panel (d). One can see in panel (c) that the transmitted radiation has two frequencies. It means that the transmitted radiation consists of two modes with different $k$, in other words the interaction of the dispersive waves with the soliton results in the partial scattering of the radiation into the second mode. The positions of the frequencies and the wavevectors of the transmitted modes are predicted by the resonant condition very well. The reflected waves contain two different modes too, the wave vectors of the reflected radiation are predicted by the resonance condition also very precisely.

The efficiency of the four-wave mixing depends on the soliton intensity controlled by the parameter $\delta_s$, as expected the scattering on more intense solitons is stronger then on the solitons of smaller intensities.

\begin{figure}[h]
\centering \epsfig{file=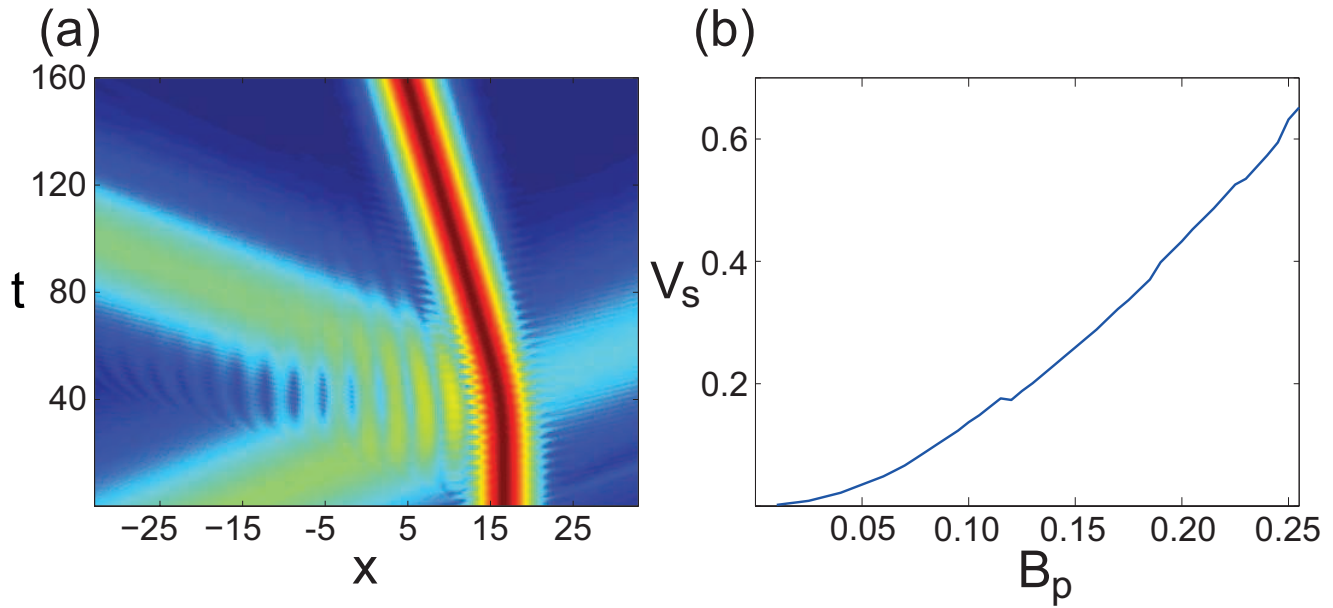, width=\textwidth}
\caption{Panel (a) shows a stable $\delta_s=0.25$ soliton interacting with a propagating plane wave injected at $x=-20$. The dependency of the  soliton velocity after the collision on the amplitude of the incident wave $B_p$ is shown in panel (b). The soliton parameter is $\delta_s=0.2$ and the wave vector of the incident waves is $k=1.5$. }
\label{fig:ZweiDreiPW}
\end{figure}

Now let us consider how the radiation affects the dynamics of the solitons. One can see that the recoil due to the four-wave mixing results in a change of the soliton velocity, see Fig.~\ref{fig:ZweiDreiPW}, where the interaction of a soliton with quasi-linear but relatively intensive wave envelope is shown. It is seen that the soliton velocity is changing in the course of the interaction. The dependence of the soliton velocity after the collision on the amplitude of the incident wave is shown in panel (b) of Fig.~\ref{fig:ZweiDreiPW}. The final velocity depends not only on the intensity but also on the duration and the frequency of the incident wave and on the intensity of the soliton.

\begin{figure}[h]
\centering \epsfig{file=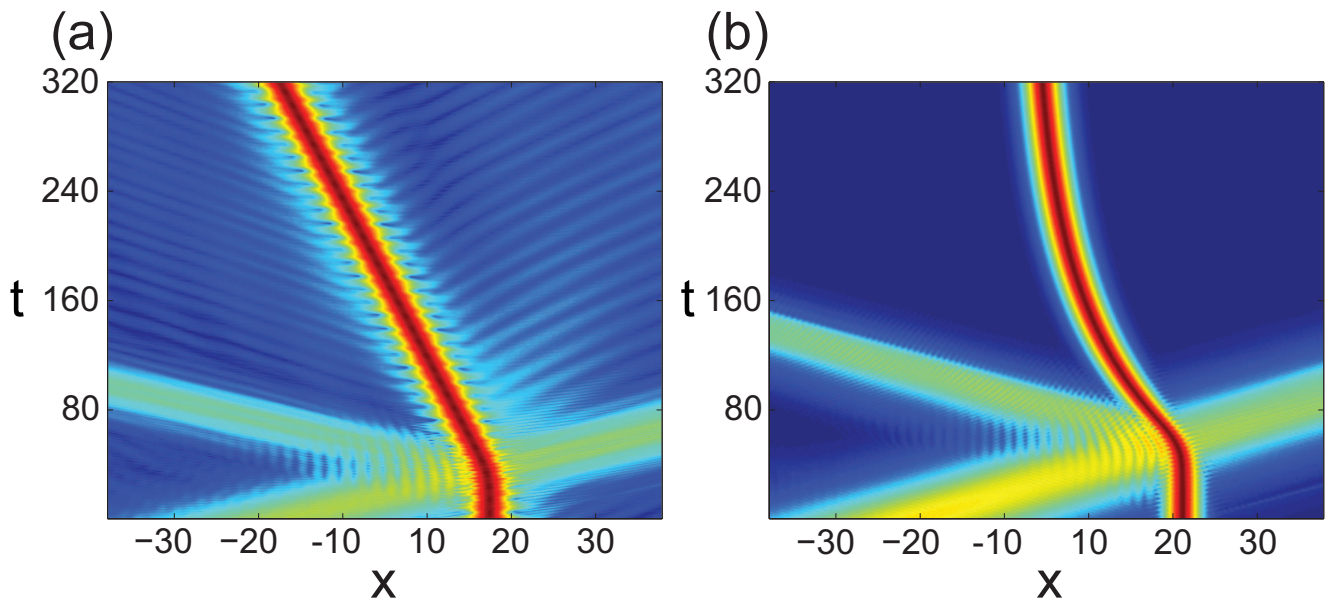, width=\textwidth}
\caption{Panel (a) shows a stationary soliton interacting with a relatively strong incident dispersive wave with $k=1.5$. The pump detuning is $\delta_c=0.25$, the intensity of the parametric pump is $\mu=0.025$, the linear losses are $\gamma=0.005$. Panel (b) shows the same for the parametric pump detuning $\delta_c=-0.1$.}
\label{fig:FigNewforPWI_2}
\end{figure}

Finally let us consider interaction between the dispersive waves and the solitons in the presence of the parametric pump. As discussed in the previous section, stable solitons can be found in the presence of pump of appropriate amplitude and frequency, an example is shown in Fig.~\ref{fig:EinPump}. We take a soliton with $\delta_s=0.2$ as the initial condition and let it propagate under the action of the parametric pump with $\delta_c=0.25$ and $\mu=0.025$. We wait until a stationary oscillating soliton forms (as in panel (a) of Fig.~\ref{fig:EinPump} ) and then irradiate it with a dispersive pulse having wave vector $k=1.5$. The collision changes the soliton velocity and the soliton continues to propagate with some \emph{non}-zero velocity emitting radiation as seen in panel (a) of Fig.~\ref{fig:FigNewforPWI_2}.

 Panel (b) illustrates the scattering of the dispersive waves on the stationary non-emitting soliton (shown in  panel (b) of Fig.~\ref{fig:EinPump}), the parameters are $\delta_c=-0.1$,  $\mu=0.025$ and $\gamma=0.005$.
The soliton is affected by an envelope of dispersive waves with $k=1.5$ leading to the bending of the soliton trajectory. The beginning seems much alike panel (a) of Fig.~\ref{fig:FigNewforPWI_2} but after the interaction with the dispersive waves the soliton returns to a resting state with no radiation wings.

\section{Soliton Interaction}

\begin{figure}[h]
\centering \epsfig{file=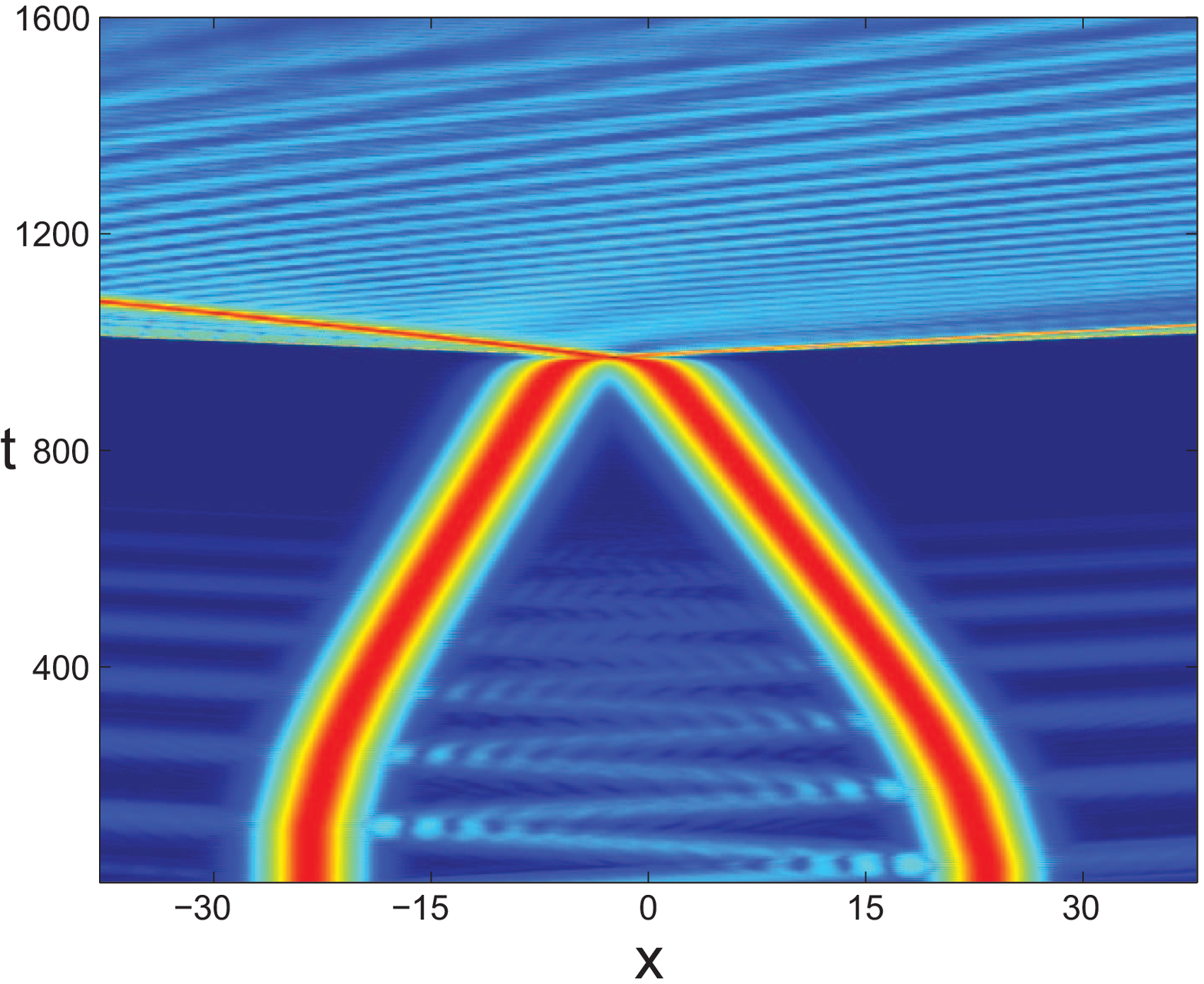, width=\textwidth}
\caption{Two stable $\delta_s=0.25$ solitons collide due to dispersive waves injected at $x=0$. The incident wave has the wave vector $k=1.0$ and the amplitude of $B=0.03$. }
\label{fig:VierPW}
\end{figure}

We start with the case without parametric pump. Then two stable solitons separated by relatively large distance (much larger comparing to the characteristic size of the solitons) do not interact and can propagate parallel to each other for extremely long distances. But the injection of dispersive waves between the solitons causes the scattering of the waves on solitons and the scattering bends the trajectories of the solitons toward each other. In Fig.~\ref{fig:VierPW} this dispersive waves mediated interaction of  two stable soliton is shown. One can see the radiation bouncing between the attracting solitons and the bending of soliton trajectories. Eventually the solitons collide and annihilate.

\begin{figure}[h]
\centering \epsfig{file=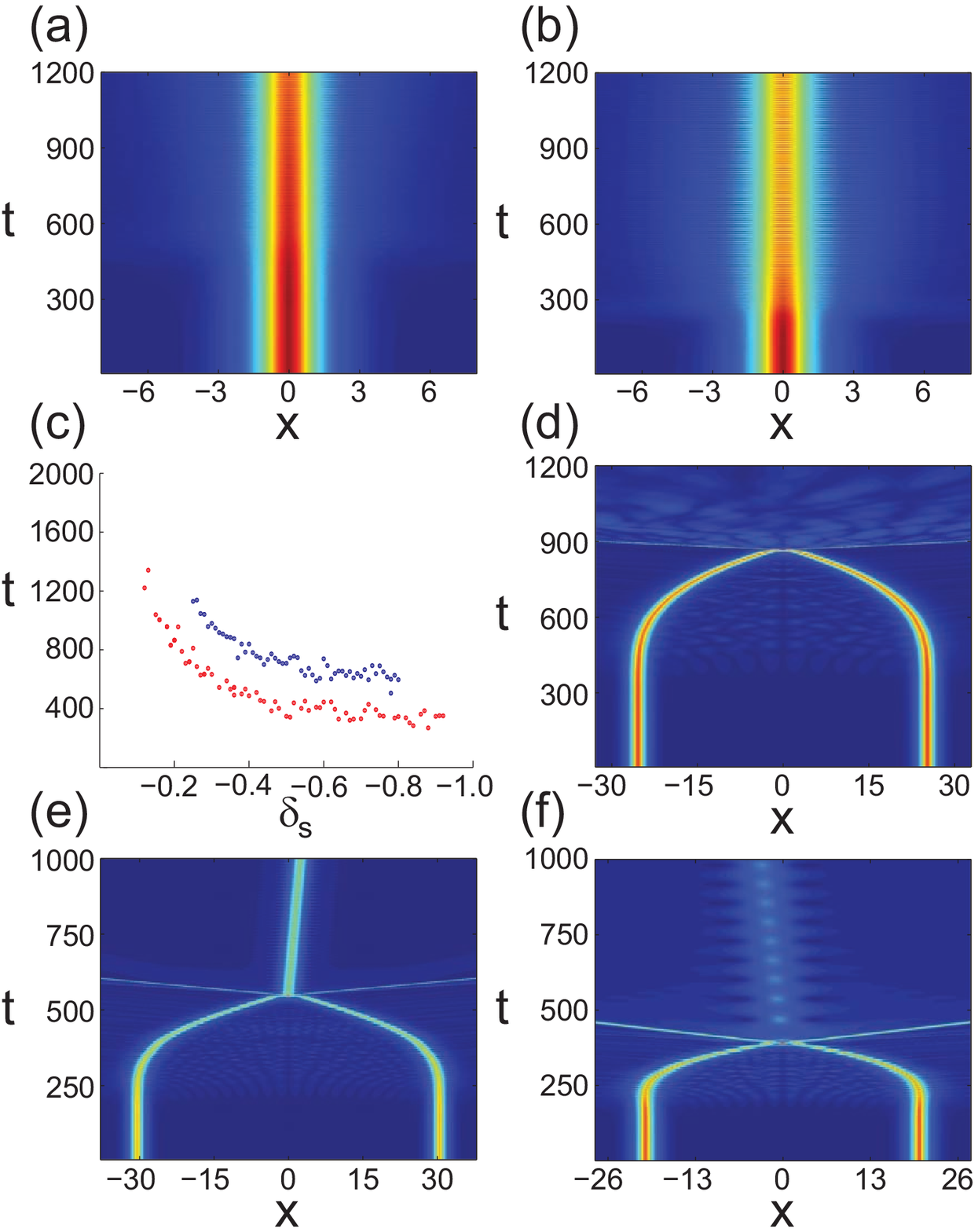, width=\textwidth}
\caption{Panel (a) shows the propagation of a radiating soliton with $\delta_s=-0.17$. Panel (b) shows the propagation of a radiating soliton with $\delta_s=-0.32$. Panel (c) shows the collision times for different $\delta_s$ for initial solitons distance $\Delta x=60$, in red, and $\Delta x=120$, in blue. Solitons separated by larger distance take longer to collide since radiation has to travel for larger inter-soliton distance. Panel (d) shows two radiating solitons with $\delta_s=-0.17$, the collision destruct both solitons. Panel (e) shows the same but for solitons with $\delta_s=-0.32$, in this case the collision generates a lower energy soliton. Panel (f) shows the behavior of two radiating solitons with $\delta_s=-0.4$, the collision results in the formation of a "breather-like" type structure.}
\label{fig:YES}
\end{figure}

In the case of radiating solitons no seeded radiation is needed for long range inter-soliton interaction, the solitons can interact through their radiation wings. In Fig.~\ref{fig:YES} different types of radiating solitons interacting through the radiation are depicted. One should expect that the solitons with larger negative $\delta_s$ will interact stronger because they emit more intense radiation. Consequently, more intense solitons should collide sooner then the solitons with low intensity, it is evidenced in panel (c) of Fig.~\ref{fig:YES} showing the collision time as a function of soliton parameter $\delta_s$. Let us note that the result of the soliton collision can be different: it can be a annihilation of soliton, the formation of one new soliton or the formation of an oscillating state, see panels (d)-(f).

\begin{figure}[h]
\centering \epsfig{file=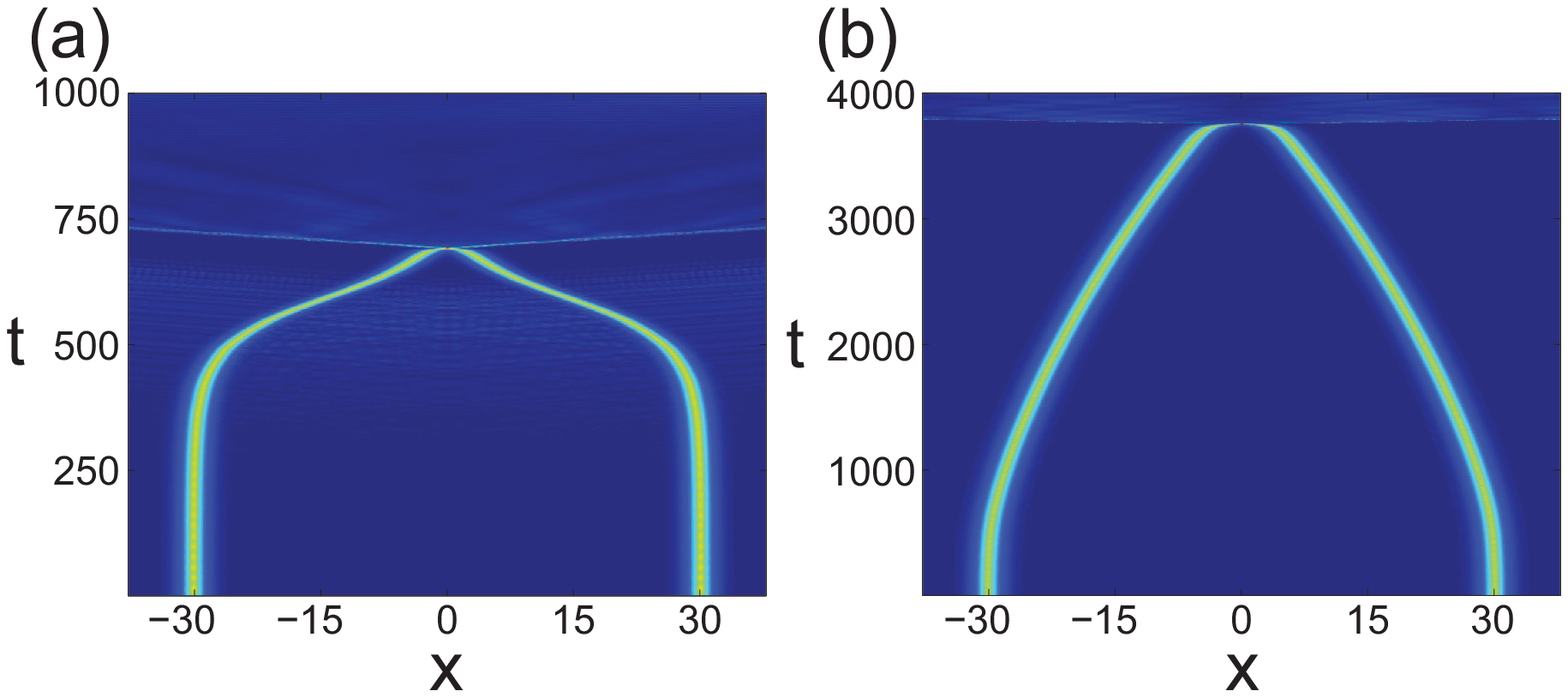, width=\textwidth}
\caption{Panel (a) shows the dynamics of radiating soliton with initial detuning $\delta_s=-0.1$ pumped by the parametric pump $\mu=0.025$ at the zero detuning $\delta_c=0$.  Panel (b) shows the collision of two soliton with the initial detuning $\delta_s=0.1$ in the presence of the pump with the detuning $\delta_c=0.36$.}
\label{fig:EinZweiPump}
\end{figure}

Finally, we consider the interactions of the solitons in the presence of parametric pump. In this case, providing that the pump has the right detuning and the sufficient intensity, the interaction can take place even if the initial condition is taken in the form of stable solitons. The pump will amplify the solitons and the intense solitons will interact though their radiation. This case is illustrated in Fig.~\ref{fig:EinZweiPump} where the interactions and the collisions of the radiating solitons are shown. Panel (a) shows the collision in the case when the initial condition is taken in the form of two unstable solitons with $\delta_s=-0.1$, the pump parameters are $\delta_c=0$ and $\mu=0.025$. In panel (b) the collision of two soliton in the presence of the parametric pump with $\delta_c=0.36$ is shown. As an initial condition stable solitons with $\delta_s=0.1$ were used. The parametric pump amplifies the solitons making their propagation constants negative, the solitons start emitting waves and the dispersive wave mediated interaction results in the eventual collision of the solitons.

\section{Motion of the solitons under the action of the dispersive waves }

Let us consider the case when the frequency of the wave launched onto the soliton is close to the frequency of the dispersive waves emitted by the soliton. Then the mutual phase $\varphi(t)$ between these waves becomes important. One can see that from the balance of conserving (without the pump and the dissipation) quantities, the energy and the momentum. Indeed, the  energy and the momentum densities of the incoming wave are known. Calculating the energy and the momentum of the wave scattered on the soliton we can find the variation of the energy and the momentum of the soliton. The change of the soliton energy and the momentum defines the change of the soliton intensity and velocity. The waves propagating outward of the soliton can be represented as a superposition of the waves scattered on the soliton and the emitted by the solton. So the interference of these wave affects the momentum and the energy densities in the waves propagating outwards the soliton. That is why the mutual phase between the soliton and the incident wave is important.

\begin{figure}[h]
\centering \epsfig{file=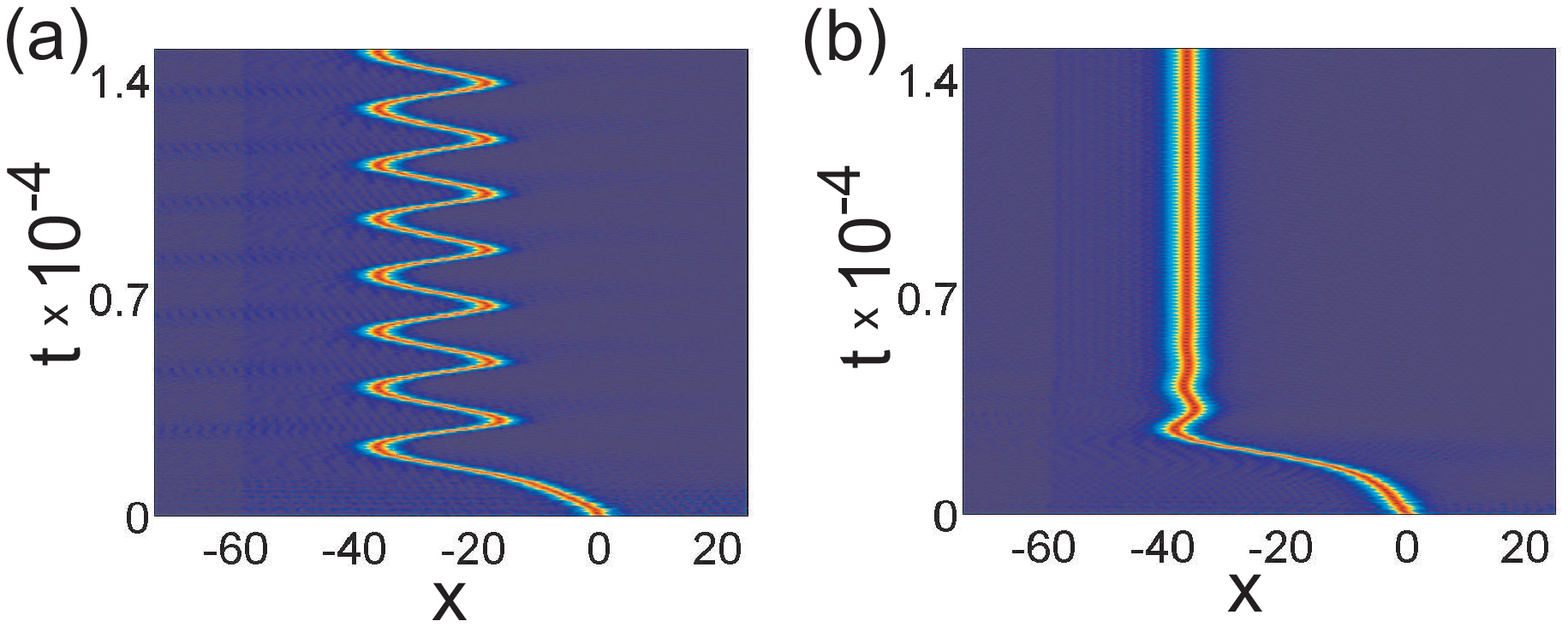, width=\textwidth}
\caption{The dynamics of the soliton under the action of the waves emitted by the source $f(x, t)=f_0 \sin(\delta_s t) \exp(-(x-x_{s0})^2/w_s^2)$. Panel (a) shows the case of the unsynchronized soliton when $\delta_s=1.8775$ and panel (b) shows the synchronized soliton when $\delta_s=1.9$. The other parameters are $f_0=0.04$,  $x_{s0}=-60$, $w_s=1$, $\mu=0.025$, $\delta_{c0}=0.05$ and $\gamma=0.005$. }
\label{fig:ext_rad}
\end{figure}

We performed the following numerical experiment. We add the spatially localized oscillating driving force $f(x, t)$  at the right hand side of the equation (\ref{e3}). This works as a source of continuous dispersive waves with fixed frequency. If we choose the frequency of the source to be close to the frequency of the waves emitted by the soliton then the soliton starts to oscillate in the field of the incident dispersive wave, see panel (a) of  Fig.~\ref{fig:ext_rad}. Changing the frequency of the source it is possible to synchronize the  soliton with the external source, the synchronization forces the soliton to relax to the stable equilibrium position, see panel (b).

\section{Radiative pinning of the solitons on inhomogeneities}

The problem we address in this section is how inhomogeneities can affect the dynamics of the radiating solitons. Indeed, if we have an inhomogeneity, for instance local variation of the refractive index, then this inhomogeneity will cause partial reflection of the waves emitted by the soliton. The reflected waves will return to the soliton and interact with it. Thus the velocity of the soliton will be changed because of the resonant scattering of the waves previously emitted by the same soliton. In this way the inhomogeneity can affect the soliton even if the distance between the soliton and the inhomogeneity is much larger than the soliton width.

\begin{figure}[h]
\centering \epsfig{file=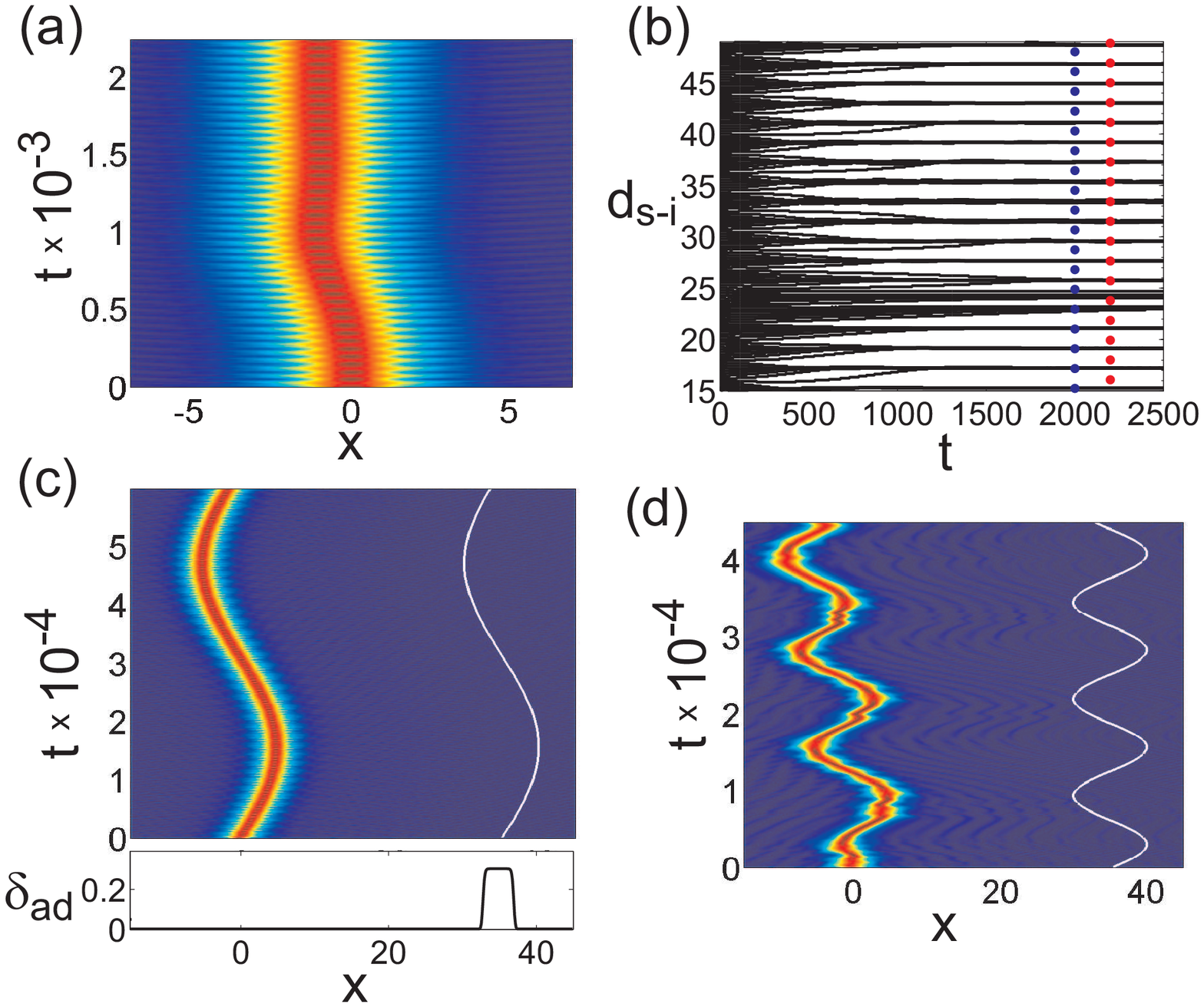, width=\textwidth}
\caption{Panel (a) shows the evolution of parametrically pumped soliton launched at the distance $d=42$ from the resting inhomogeneity. In panel (b) the trajectories of the solitons launched at different distances from the resting inhomogeneity are shown. The blue and red dots show the predicted equilibrium distances, see text for more details. Panels (c) and (d) show the dynamics of the soliton when the position of the inhomogeneity oscillate in time with the frequencies $\delta_d=0.0001$ and $\delta_d=0.0005$ correspondingly. The profile of the inhomogeneity $\delta_d=\delta_{ad}sec((x-x_{d})^8/w_d^8)$  is shown in the lower part of panel (c). The trajectory of the centre of the inhomogeneity $x_d=x_{d0}+x_{ad}\sin(\delta_d t)$ is shown by the white curve.  The parameters are  $\delta_{ad}=0.05$, $w_d=2$, $x_{ad}=5$, $\mu=0.025$, $\delta_{c}=0.05$ and $\gamma=0.005$.}
\label{fig:along_the wall}
\end{figure}

It is interesting that if the wave reflected from the inhomogeneity returns to the soliton at the proper phase then this wave does not affect the velocity of the soliton. It means that the soliton launched at a particular distance from the inhomogeneity will  get locked and keep the same distance from the inhomogeneity. Panel (a) of  Fig.~\ref{fig:along_the wall} shows how the soliton moves and get pinned at the equilibrium position under the action of the waves reflected from the inhomogeneity.

Since the described effect depends on the phase of the reflected wave the pinning distances must be periodic given by the ratio
\begin{align}
k d_{en}=\pi n+\varphi_0, \label{eq_dist}
\end{align}
where $d_{en}$ is the equilibrium distance between the soliton and the inhomogeneity, $k$ is the wavenumber of the mediating wave, $n$ is an integer and $\varphi_0$ is a constant to be found independently. This is illustrated in  panel (b) of Fig.~\ref{fig:along_the wall}. The trajectories of the solitons launched at different distances from the inhomogeneity are shown in panel (b). It is seen that after some transitional processes the solitons select and retain the position from a discrete set of the equilibrium distances from the inhomogeneity.

The locked states are stable and survive even if the position of the inhomogeneity varies adiabatically: then the position of the soliton varies too keeping the distance between the soliton and the inhomogeneity constant. This process is illustrated in panel (c) of Fig.~\ref{fig:along_the wall}.  So the solitons stay pinned on the inhomogeneity. The peculiarity of this pinning is that it is mediated by radiative waves and so can take place at the distances much larger than the width of the soliton when the soliton does not feel the inhomogeneity directly. In the presence of the linear losses the distance must be not too large otherwise the radiated waves will simply decay and there will be no reflected wave affecting the soliton. We notice that when the position of the inhomogeneity varies fast then the soliton cannot follow it and start drifting, see panel (d).

To prove that the pinning happens because of the interaction mediated by the soliton emitted waves we extracted the wave number from the radiation spectrum of the soliton (the first sideband, see panel (d) of Fig.~\ref{fig:instab}, the curve for $\mu=0.025$) and mark the equilibrium distances given by (\ref{eq_dist}) by the dots. The difference between the blue and red dots is that the blue dots are shifted to give a perfect match for the equilibrium distance $d_{en} \approx 15$ and the red ones to give the perfect match for the equilibrium distance $d_{en} \approx 45$. One can see that the predicted and the observed positions match very well. However a breakup of periodicity was observed at around $d_s \approx 25$, probably because the mediating waves in reality have many spectral lines.

\section{Conclusion}

We have shown that the parametric pump can result in the formation of stationary gap solitons emitting dispersive waves. In numerical simulations we have observed the formation of the stable solitons emitting dispersive waves if the resonant frequency of the parametric pump lies in the gap of the dispersion characteristics of the linear waves. This pump does not amplify small excitations because their frequencies cannot be in the gap and so these excitations are detuned from the parametric resonance. In the same time the frequency of the soliton is shifted because of nonlinear effect and lies in the gap. So the energy can be delivered to the solitons by the parametric gain. It allows to compensate for the radiative loss and make the emitting solitons stationary.

The analysis of the spectral properties of the soliton radiation  revealed that the effect responsible for the emission of dispersive waves by the stationary solitary structures in the presence of parametric gain is same as the effect leading to the instability of gap solitons in the absence of parametric pump. It was found that if the parametric pump is weak then the spectral characteristics of the emitted radiation matches well the spectrum of the mode generating the instability of the conservative gap solitons. We claim that the formation of radiating gap solitons can be understood as an interplay of two effects. The first effect is the emission of radiation because of resonant interaction between the linear waves and the soliton. The second effect is amplification of the soliton by the parametric gain.

The interaction of the solitons with the dispersive waves was studied. We derived the condition of the resonant four-wave mixing between the gap solitons and the dispersive waves. By direct numerical simulations we investigated the scattering of the envelopes of the dispersive waves on gap solitons and found that the resonance conditions predict the wave vectors and the frequencies of the scattered waves very well. We also demonstrated that gap solitons separated by the distance much larger than the soliton length can efficiently interact through the exchange of the dispersive waves. This interaction results in the mutual attraction and collision of the solitons, after the collision the solitons can annihilate or produce a new soliton or an oscillating localized state.

The interactions of a radiating soliton with dispersive waves were also investigated.  It is shown that if the frequency of the incident dispersive wave is close to the frequency of the soliton emitted radiation then the mutual phase between the soliton and the incident radiation becomes important. It is shown that the soliton can oscillate in the field of the continuous incident wave or be synchronized with it.

The case when the emitting solitons are excited in the cavity with spatial inhomogeneity is also studied in the paper. It was shown that the wave reflected from the nonlinearity can interact with the soliton and that this interaction selects an equilibrium distance between the soliton and the inhomogeneity. So, because of the dispersive waves mediated interaction, the solitons can be pinned to an inhomogeneity even if the distance between the soliton and the inhomogeneity is much larger then the soliton length.

\bigskip

\section*{Acknowledgements}
AVY and LRG acknowledge support of the FCT (Portugal) under the grants  PTDC/FIS/112624/2009, and PEst-OE/FIS/UI0618/2011, KS acknowledges financial support by spanish Ministerio de Education y Ciencia and European FEDER through project FIS2011-29734-C02-01.

\end{document}